\documentclass[12pt]{iopart}

\usepackage[utf8]{inputenc}
\usepackage[T1]{fontenc}
\usepackage{amssymb}
\usepackage{graphicx}
\usepackage{hyperref}
\usepackage{color}

\usepackage{booktabs} 
\usepackage{subcaption}

\usepackage{algorithm}
\usepackage{algorithmicx}
\usepackage{algpseudocode}

\begin{document}

\title{Dynamical analysis of financial stocks network: improving forecasting using network properties}

\author{Ixandra Achitouv}
\ead{ixandra.achitouv@cnrs.fr}
\address{Institut des Syst\`emes Complexes ISC-PIF , CNRS, 113 rue Nationale,
Paris, 75013, France.}

\date{\today}

\begin{abstract}
Applying a network analysis to stock return correlations, we study the dynamical properties of the network and how they correlate with the market return, finding meaningful variables that partially capture the complex dynamical processes of stock interactions and the market structure. We then use the individual properties of stocks within the network along with the global ones, to find correlations with the future returns of individual S\&P 500 stocks. Applying these properties as input variables for forecasting, we find a 50$\%$ improvement on the R2score in the prediction of stock returns on long time scales (per year), and 3$\%$ on short time scales (2 days), relative to baseline models without network variables. 
\end{abstract}

%\maketitle
%\subsection*{ }
%Keywords: Complex network,  Stock returns correlation, Stock market, Random Matrix Theory

\section{Introduction}
\label{introduction}

Financial market dynamics exhibit properties of complex systems, such as non-linear behaviour and inter-dependencies of stock prices \cite{mantegna1999hierarchical}\cite{Bouchaud2003} \cite{Cont2001}, emergence phenomena such as trends and bubbles from interactions of market participants \cite{Farmer1999}, \cite{Lux1999}, \cite{Sornette2003}, adaptive behaviour \cite{Arthur1997}, \cite{Lo2004} and feedback mechanisms \cite{LeBaron2001} (e.g. rising prices attract more investors, further driving up prices). Collective behaviours emerging in financial markets are known as market modes, referring to synchronized movements of groups of stocks. They have been studied in stock return correlations using different tools including principal component analysis (e.g. \cite{jolliffe2002principal}), complex network analysis (e.g. \cite{mantegna1999hierarchical},\cite{boccaletti2006complex},\cite{peron2011collectivebehaviorfinancialmarket},\cite{achitouv2024inferringfinancialstockreturns}) and random matrix theory (e.g. \cite{laloux1999noise},\cite{plerou1999universal},\cite{bouchaud2009financial}). Indeed, in a complex system, the behavior of individual components, or agents, is coupled to the collective dynamics of the system. When applied to financial markets, one could ask how these couplings can be extracted and used to improve forecasting of the global stock market as well as in individual stock returns.

Many studies have been proposed to forecast individual stock returns using soft computing methods \cite{yang2002stock}, \cite{kim2005financial}, \cite{kim2005financial} (e.g. machine learning regression algorithms) using mostly individual features of a stock (e.g. previous closing value, difference between high and low value of the stock each day, mean volume exchanged, etc..). Others have included macroeconomic variables to forecast stock returns (e.g. \cite{li2014forecasting}), or sentiment analysis using natural language processing techniques  (e.g. \cite{bollen2011twitter}, \cite{schumaker2009textual}, \cite{plakandaras2015market}). Some studies have analyzed stock market networks' structural properties to detect influencers in stock market ``communities" (e.g. \cite{namaki2011network} \cite{huang2009cumulative}) but little research has been conducted to apply the properties of a network built on stock returns correlations to forecast the future movements of the global stock market \cite{kim2016predicting}, \cite{SEONG2022107608}, or of individual stocks \cite{baitinger2021forecasting}. 

In this article we use complex network analysis to study the dynamical inter-dependencies of stock returns and their collective behaviours, and test how the properties of the network correlate with the market stock return that we define as the unweighted average return across all assets. Indeed, networks underpin complex systems and influence their behavior\cite{newman2010networks}. Therefore, by analysing the dynamical evolution of the stock returns network one could hope to infer the overall dynamics of the stock returns. We further consider the properties of individual stocks within the network and test how these properties correlate with the future returns of individual stocks. Finally, we consider how our results change depending on the time scale employed to built the network, to test if we have a scale invariance on the correlations observed between the log return and the network properties. Indeed, scale-invariance is often considered a property of complex systems \cite{stanley1996anomalous}, including in financial market \cite{calvet2002multifractality}, \cite{stanley1996anomalous}. The fractal market hypothesis, discussed in \cite{mandelbrot1982fractal}, \cite{peters1994fractal}, \cite{mandelbrot1997fractals}, \cite{mandelbrot2004misbehavior} suggests that market prices exhibit fractal properties and that patterns observed in short-term price movements can resemble those in long-term trends. This contrasts with the efficient market hypothesis, which assumes that market prices follow a random walk and are inherently unpredictable. Thus it becomes interesting to test the added value of the network input variable on short and long time scale forecasting and study if the key predictors are the same.

This article is organized as follows: in sec.\ref{secnetwork} we describe the main network properties we consider, and how we built our financial network. In sec.\ref{secglobal} we measure the dynamical evolution of the network properties and test for its correlation with the overall market return. In sec.\ref{secindiv} we consider individual properties of the stocks within the network and test if they carry meaningful correlations with their future returns. We then leverage these properties to forecast individual stock returns. In sec.\ref{secconclu} we present our conclusion.

\section{Network properties and construction of the financial stock return network}\label{secnetwork}

\subsection{Network properties}
In order to study the properties of the network we consider centrality measures:\\

(1) The degree $k_i$ of a node $i$ in a network represents the number of connections it has to other nodes\cite{newman2010networks}. In an undirected network, the degree $k_i$ of a node $i$ is given by the sum of the adjacency matrix elements corresponding to that node.

\begin{equation}
    k_i= \sum_{j} a_{ij}
\end{equation}

where $a_{ij}$ is an element of the adjacency matrix A, which is 1 if there is an edge between nodes $i$ and $j$, and 0 otherwise. \\

(2) Closeness centrality $C_{C}(u)$  of a node $u$ measures the average length of the shortest path from a node to all other nodes in the network, representing how quickly information spreads from a given node to others \cite{freeman1978centrality}. It identifies nodes that can quickly interact with all other nodes, and is defined as:

\begin{equation}
    C_{C}(u)=\frac{1}{\sum_{v\in V} d(u,v)} 
\end{equation}

where $d(u,v)$ is the shortest path distance between nodes $u$ and $v$, and V is the set of all nodes in the network.\\

(3) Betweenness centrality $C_B(v)$ of a node $v$ is a measure of the extent to which a node lies on the shortest paths between other nodes, indicating its role as a bridge or connector within the network. Nodes with high betweenness centrality can control the flow of information or resources between different parts of the network \cite{freeman1977set}.

\begin{equation}
    C_B(v)=\sum_{v\neq s\neq t} \frac{\sigma_{st}(v)}{\sigma_{st}}
\end{equation}
where $\sigma_{st}$ is the total number of shortest paths from node $s$ to node $t$ and $\sigma_{st}(v)$ is the number of those paths that pass through $v$. \\

(4) Eigenvector centrality $C_E(i)$ of a node $i$ is a measure of the influence of a node in a network based on the idea that connections to high-scoring nodes contribute more to the score of the node in question\cite{bonacich1987power}. A high eigenvector score means that a node is connected to many nodes who themselves have high scores. It is computed using the principal eigenvector of the adjacency matrix of the network:

\begin{equation}
    C_E(i)=\frac{1}{\lambda}\sum_{j\in \mathcal{N}(i)}A_{ij}C_E(j)
\end{equation}
where $\lambda$ is a constant (the eigenvalue), $\mathcal{N}(i)$ is the set of neighbors of $i$, and A is the adjacency matrix of the network. Applied to a financial market, high eigenvector values can be problematic for systemic risks as contagion risk is linked to these influential stocks \cite{zhu2016measuring}. Indeed, a highly interconnected financial network can act as an amplification mechanism by creating channels for
a shock to spread, leading to losses that are much larger than the initial changes \cite{jackson2020systemicriskfinancialnetworks}.\\

In addition, we consider the weighted clustering coefficient $C_C(i)$ of a node $i$, it measures the tendency of nodes to cluster together while taking into account the strength of the connections (weights) between nodes, providing a more nuanced view of network cohesiveness\cite{barrat2004architecture}: 

\begin{equation}
    C_C(i)=\frac{1}{s_i(k_i-1)}\sum_{j,h}\frac{(w_{ij}+w_{ih})}{2}a_{ij}a_{ih}a_{jh}
\end{equation}
where $s_i$ is the sum of the weights of the edges connected to node $i$, $k_i$ is the degree of node $i$, $w_{ij}$ is the weight of the edge between nodes $i$ and $j$, and $a_{ij}$ is the adjacency matrix element that is 1 if nodes $i$ and $j$ are connected and 0 otherwise. Applied to a financial market, networks with high clustering coefficients are often more robust and resilient to node failures, as the redundancy in connections within clusters can help maintain the overall connectivity \cite{albert2000error}. This could be used as a network systemic risk measure similarly to the eigenvector centrality, when building on optimal portfolio. In fact it was found (e.g. \cite{heshmati2021portfolio}) that one can improve on traditional portfolio  selection by maximizing the usual Sharpe ratio where the standard deviation of the portfolio is replaced by a weighted average of node centrality measures (e.g. eigenvector centrality, betweennness centrality, closeness centrality). \\

We also measure properties that characterize the network structure globally, such as:\\

(1) Community stability measure. In a network, this refers to the persistence of community structures over time or under perturbations, indicating how robust the communities are to changes in the network. It can be quantitatively assessed using various metrics, one of which involves evaluating the modularity Q of the network \cite{newman2006modularity}. The modularity of a network is defined as:

\begin{equation}
    Q=\frac{1}{2m}\sum_{ij}(A_{ij}-\frac{k_ik_j}{2m})\delta(c_i,c_j)
\end{equation}
where $A_{ij}$ is the adjacency matrix, $k_i$ and $k_j$ are the degrees of nodes $i$ and $j$ respectively, $m$ is the total number of edges, and $\delta(c_i,c_j)$ is a function that is 1 if nodes $i$ and $j$ are in the same community and 0 otherwise.\\

(2) The largest component L, defined as the largest subset of nodes such that there is a path connecting any two nodes within this subset. It represents the most extensive cluster of interconnected nodes in the network, providing insights into the network's structure and functionality\cite{newman2010networks}. Let G=(V,E) be a graph where V is the set of vertices and E is the set of edges. If $C_i$ represents the $i$-th connected component of G, then the largest component L is given by:
\begin{equation}
    L=max_{i}\mid C_i\mid
\end{equation}
where $\mid C_i\mid$ is the size (number of nodes) of component $C_i$. 

(3) The resilience of a network, R, measures its ability to maintain its structural integrity and functionality in the face of failures or attacks on its nodes or edges. It characterizes the robustness and vulnerability of networks \cite{newman2010networks}\cite{albert2000error}. Resilience can be quantified by assessing the size of the largest connected component after a certain fraction of nodes or edges have been removed.

\begin{equation}
    R(f)=\frac{\mid L(f)\mid}{\mid V\mid}
\end{equation}
where $\mid L(f)\mid$ is the size of the largest connected component after removing $f \mid V\mid$ nodes, and $\mid V \mid$ is the total number of nodes in the original network.

\subsection{Construction of the network}

\begin{figure}
    \centering
    \begin{minipage}[b]{0.45\textwidth}
        \centering
        \includegraphics[width=\textwidth]{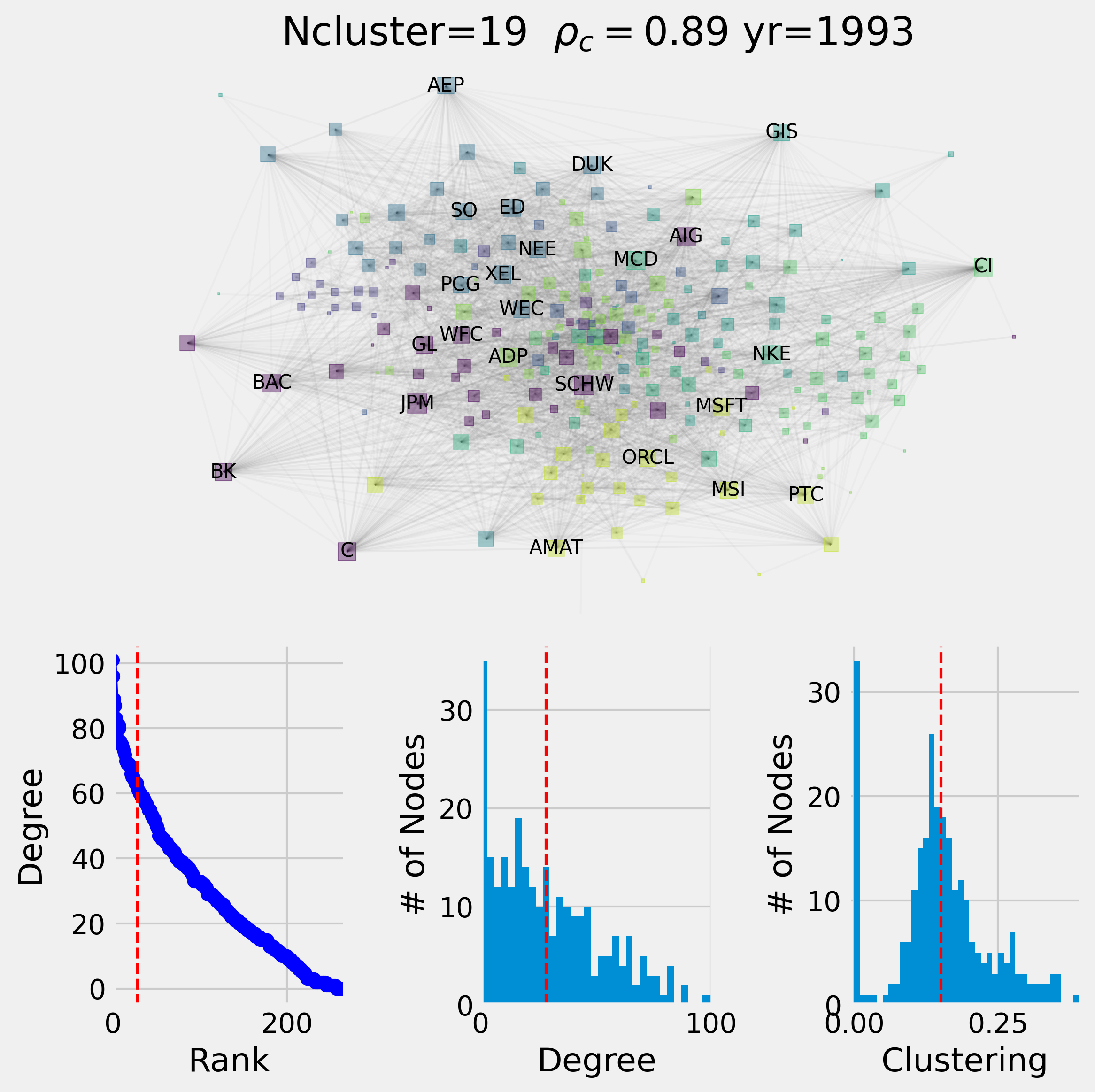}
        %\caption{Caption for Figure 1}
       % \label{fig:1}
    \end{minipage}
    \hfill
    \begin{minipage}[b]{0.45\textwidth}
        \centering
        \includegraphics[width=\textwidth]{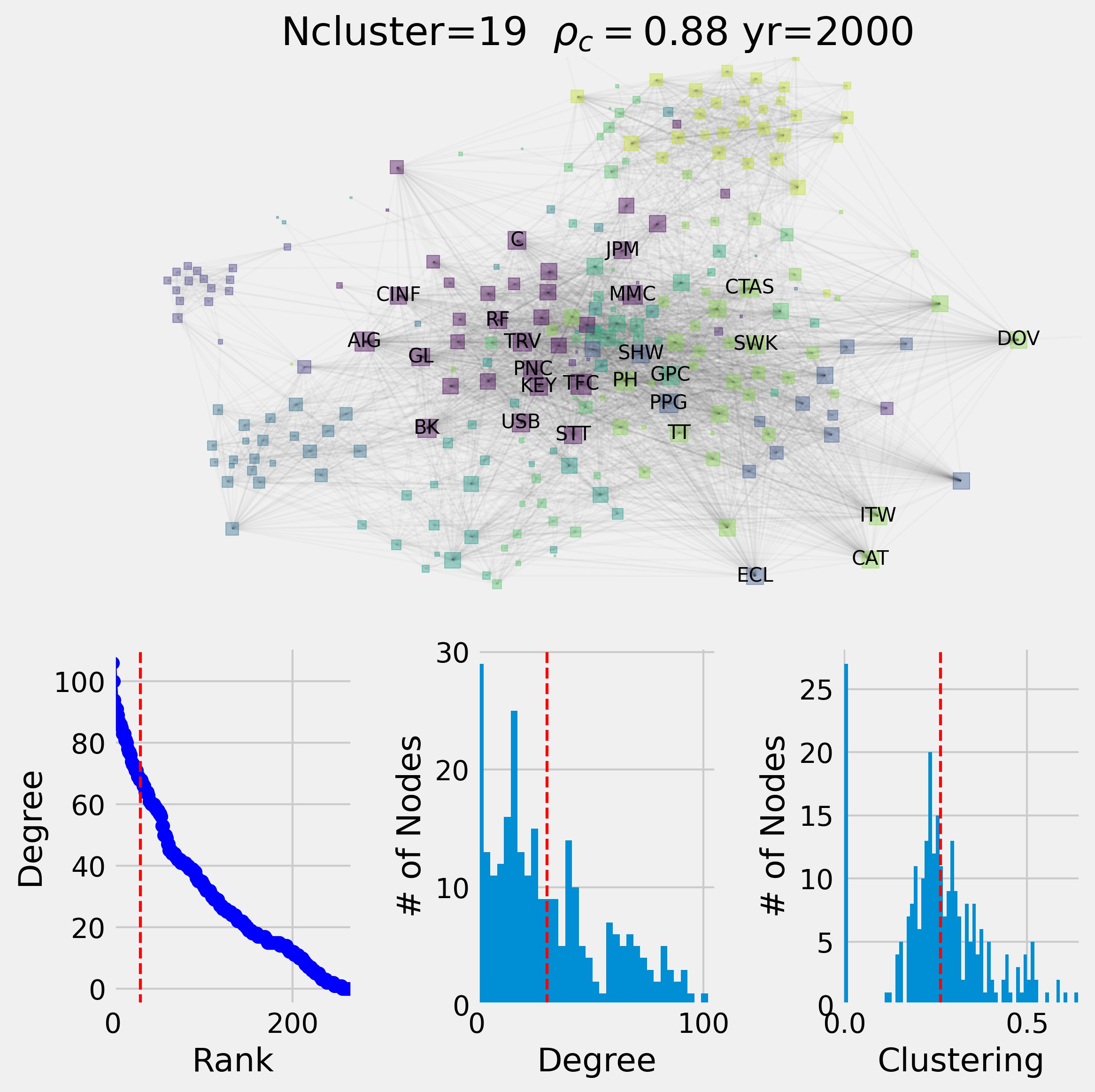}
        %\caption{Caption for Figure 2}
        %\label{fig:2}
    \end{minipage}
    \vfill
    \begin{minipage}[b]{0.45\textwidth}
        \centering
        \includegraphics[width=\textwidth]{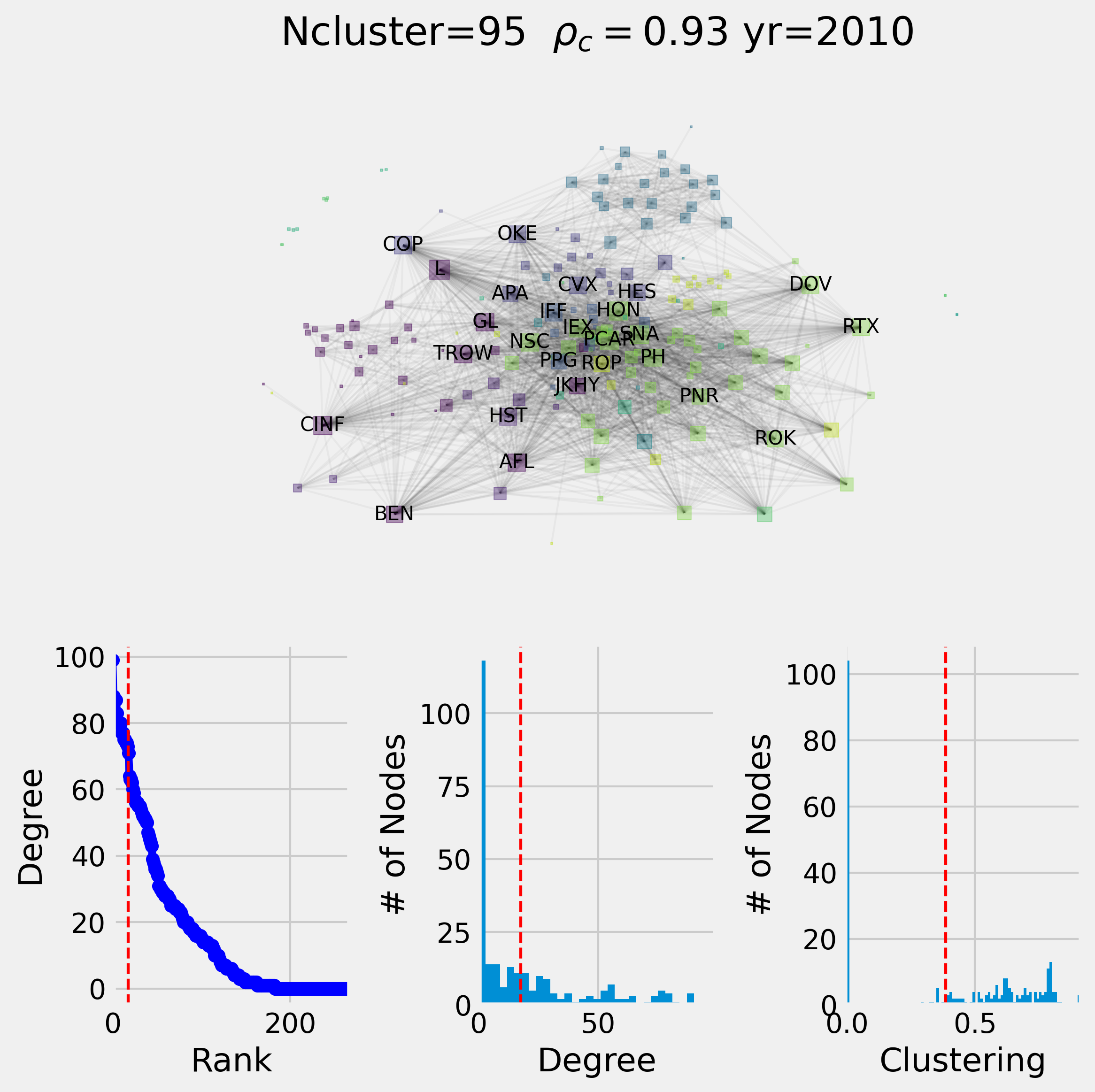}
       % \caption{Caption for Figure 3}
        %\label{fig:3}
    \end{minipage}
    \hfill
    \begin{minipage}[b]{0.45\textwidth}
        \centering
        \includegraphics[width=\textwidth]{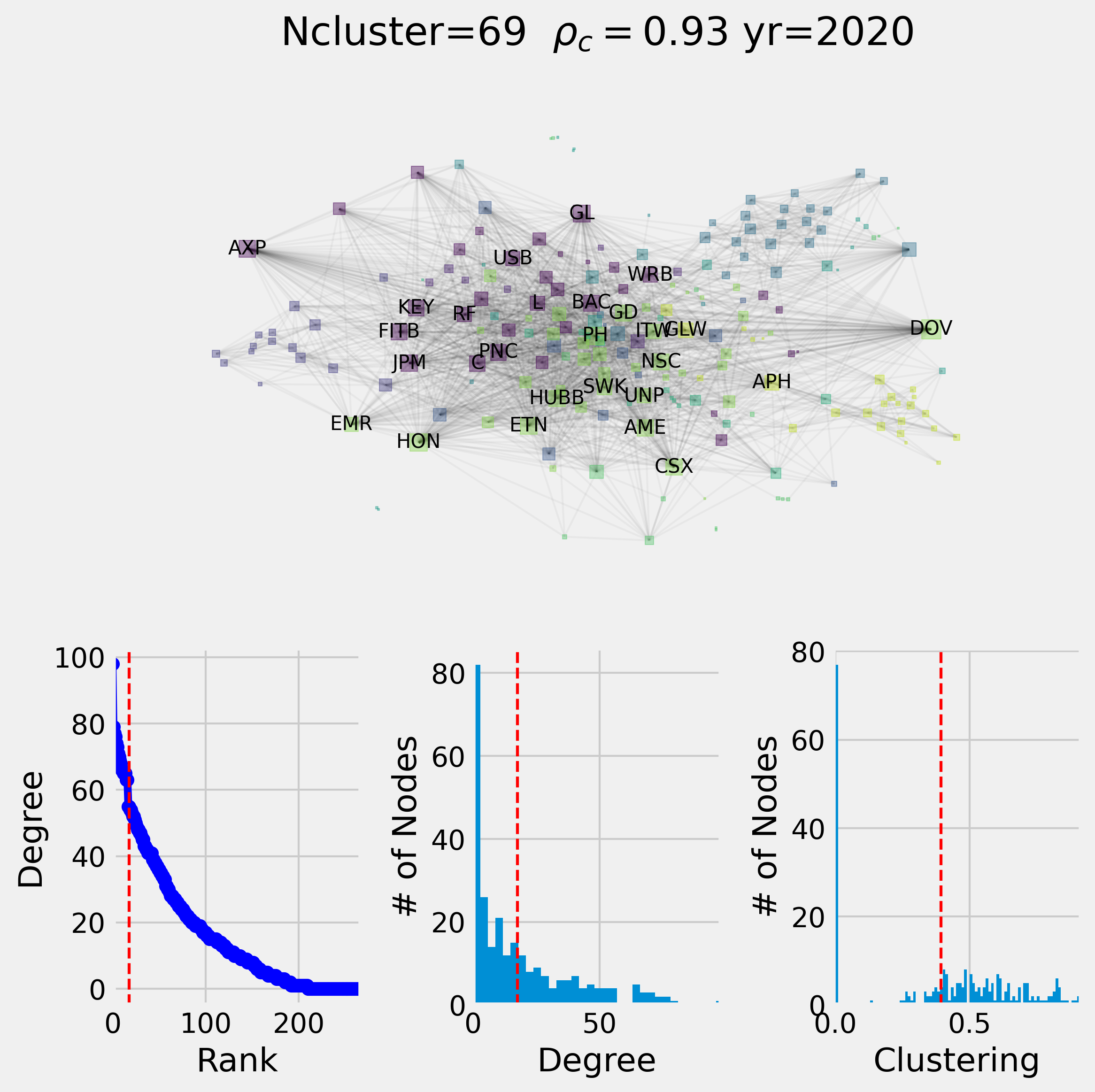}
       % \caption{Caption for Figure 4}
        %\label{fig:4}
    \end{minipage}
    \caption{Network of the stocks for different years (top panels) along with the degree distribution (left bottom panels), eigenvector centrality histogram (middle bottom panels) and clustering histogram of the stocks (right bottom panels). Red vertical lines correspond to the mean.}
    \label{fig:network_sample}
\end{figure}

For the long term period, we consider the daily stocks closing price from the S$\&$P500 ranging from 1993-01-01 to 2024-01-01 downloaded from $Yahoo$ finance. 
We clear out stocks that were not present in the entire time range, ending up with 267 stocks and 7805 days of closing values for each stock. We then split our data into 30 yearly samples covering $Nobs=30$ years of business days. 

For the short term period, we consider hourly stocks prices from the S$\&$P500, from 2022-08-31 to 2024-07-31, downloaded from $Yahoo$ finance\footnote{this time interval corresponds to the freely available data we could download at the time of this analysis.}. 
We clear out stocks that were not present in the entire time range, ending up with 488 stocks and 3346 hours of recorded price for each stock. We then split our data in time intervals of 14 hours (corresponding to two business days) and we end up with $Nobs=$238 measures (one measure every two business days).

The structure of the financial market can be mapped as a network where nodes represent the stocks and the edges connecting them represent the correlations between their returns after applying some filtering, e.g. \cite{boginski2005statistical}, \cite{onnela2004clustering}, \cite{kim2002weighted}, \cite{mantegna1999hierarchical}. 

To compute the correlation of their returns we take the closing price $P_i(t)$ log-return:
\begin{equation}
    C_{i,j}=\frac{< r_i(t) r_j(t)>-<r_i(t)><r_j(t)>}{\sigma_i\sigma_j}\label{eqCij}
\end{equation}
where $r_i(t)=\log[P_i(t)]-\log[P_i(t-1)]$ and $\sigma_i$ is the standard deviation of the stock computed over the year we consider. 

In this correlation matrix all stocks are connected to one another (the degree distribution is therefore a constant equal to the number of nodes), which does not provide any useful information. One approach to convert this correlation matrix to an adjacency matrix $A_{i,j}$, is the threshold method originally introduced in \cite{boginski2005statistical} and used in many studies e.g. \cite{namaki2011network}, \cite{huang2009cumulative}, \cite{xu2018efficient}. It is defined as $A_{i,j}=C_{i,j}$ if $|C_{i,j}| \ge \rho_{c}$, otherwise $A_{i,j}=0$. One can easily understand what it does  qualitatively: it filters out 'spurious' correlations and keeps an edge between stocks when the correlation of their return is above a certain limit. Thus remains the question of which value one should take for the threshold to optimally filter the correlation matrix of the returns, differentiating between noise and signal\footnote{Note that one could start by extracting the market modes from the correlation matrix as in \cite{achitouv2024inferringfinancialstockreturns} before applying the threshold, however this requires the correlation matrix to be full rank which may not be the case for short time scales.}. In previous studies, the value for the threshold is usually an heuristic choice as discussed in \cite{park_perspective_2020}. For instance, in \cite{nobi2014effects}, $\rho_c=<C_{i,j}>+n\sigma$, where $\sigma$ is the standard deviation of the $C_{i,j}$ distribution and $n$ an integer. Recently, a criterion based on network properties was introduce in \cite{Xu_2018}. In our work we also propose to privilege a value based on the network. Many complex systems naturally evolve towards a scale-free structure due to the dynamic processes governing their formation. The scale-free property means that the degree distribution follows a power law: a few nodes are connected with many (high degree) while most are not connected (null degree). The scale-free network emerges because the probability of a node gaining new connections is proportional to its existing connectivity \cite{barabasi2003scale}. In fact, many natural networks grow over time by the addition of new nodes. New nodes are more likely to attach to existing nodes that already have a high degree of connectivity, leading to a few highly connected hubs. This process, known as preferential attachment, is a fundamental mechanism in the formation of scale-free networks. Scale-free networks may also provide an evolutionary advantage by enabling complex systems to adapt and evolve more efficiently. Highly connected nodes (hubs) can serve as control points that help the network respond to changes and perturbations, thus enhancing the system's overall adaptability and resilience \cite{barabasi2004network}. Hence we propose to define the threshold value as the minimum value where the degree distribution of the nodes follows a power law. In practice we increase the threshold until the degree distribution becomes convex, leading to values $\rho_c \sim 0.9$. Once we have built the adjacency matrix, we use the $NetworkX$ library \cite{hagberg2008exploring} to visualize the network and compute most of the network statistics we previously introduced.

In Fig.\ref{fig:network_sample} we display the resulting networks for 4 different years, where nodes are colored by the sector of the stocks. The spatial visualisation of the network is computed using ForceAtlas2 \cite{jacomy2014forceatlas2} which maximizes/minimizes the distance of nodes that have low/high weighted edges respectively. The size of the nodes is proportional to their degree. Interestingly we can observe some clustering where stocks of the same sector (shown by colors) are closer. We run a Louvain community finder \cite{blondel2008fast} which optimizes locally  the difference between the number of edges between nodes in a community and the expected number of such edges in a random graph with the same degree sequence. 
As a result, we identify different numbers of clusters depending on the year (Ncluster). For visibility we display the label of the nodes if the node is in the top 3$\%$ of the eigenvector centrality distribution, when its eigenvector centrality is the maximum within the Ncluster, or when it has the largest eigenvector centrality within its S\&P5500 sector. 

In the lower panels of Fig.\ref{fig:network_sample} we display the degree distribution of the nodes as well as the histograms of eigenvector centrality and local clustering. The red vertical line corresponds to the mean. In the appendix Fig.\ref{fig:network_sampleh} we also show 4 selected random networks built using the short time period.

\section{Global evolution of the network and the market stock returns}\label{secglobal}

To characterize the global dynamical evolution of the network we consider the following measures: the average degree of the top $90\%$ nodes, the mean closeness centrality (the mean corresponds to the average over all nodes), the mean betweeness centrality, the mean eigenvector centrality, the mean clustering, the largest component, the resilience of the network, and the community stability. We also measure the maximum eigenvalues of stock returns computed from their correlation matrix, as this can used as a precursor to financial crises by indicating market turbulence and systemic risk \cite{laloux1999noise}.

\subsection{Long term period}

\begin{figure}
\centering
\includegraphics[width=1\linewidth]{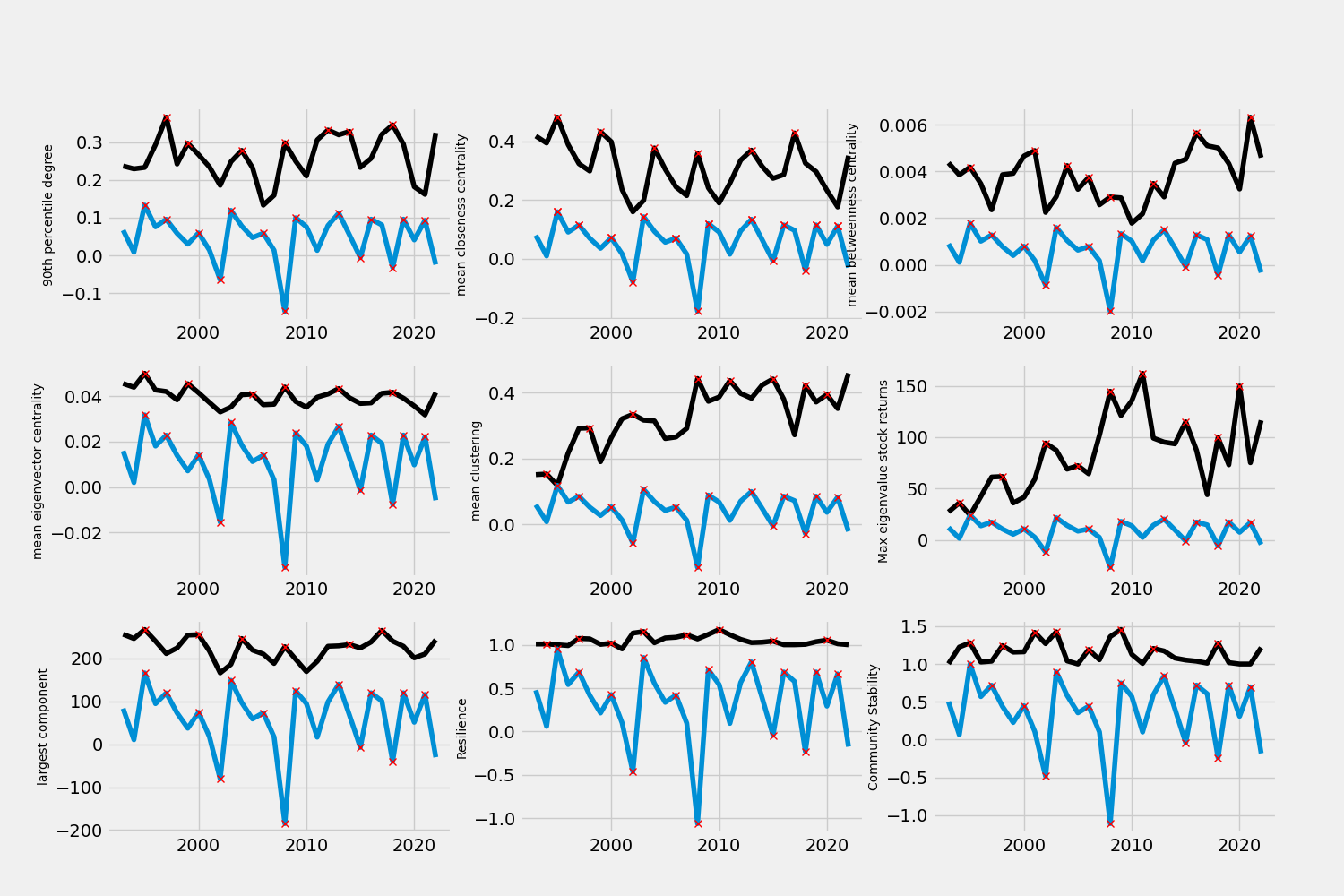}
\caption{Evolution of the network properties as a function of the year (black curves). The blue curves correspond to the stock return rescaled. The number of observed time scale is $Nobs=30$. The red crosses correspond to peaks on the curves.}\label{fig:dyna_network}
\end{figure}

In fig.\ref{fig:dyna_network} we display these measurements (black curves) along with the global stock returns evolution (blue curves) rescaled by some constant factor to fit on the same y-axes. The red crosses correspond to peaks in the curves. Interestingly it seems that qualitatively the mean clustering and the largest eigenvalue of the stock returns increase over the last 30 years which would indicate that the market becomes more connected (largest eigenvalue is the market mode).

\begin{figure}
    \centering
    \includegraphics[width=1\linewidth]{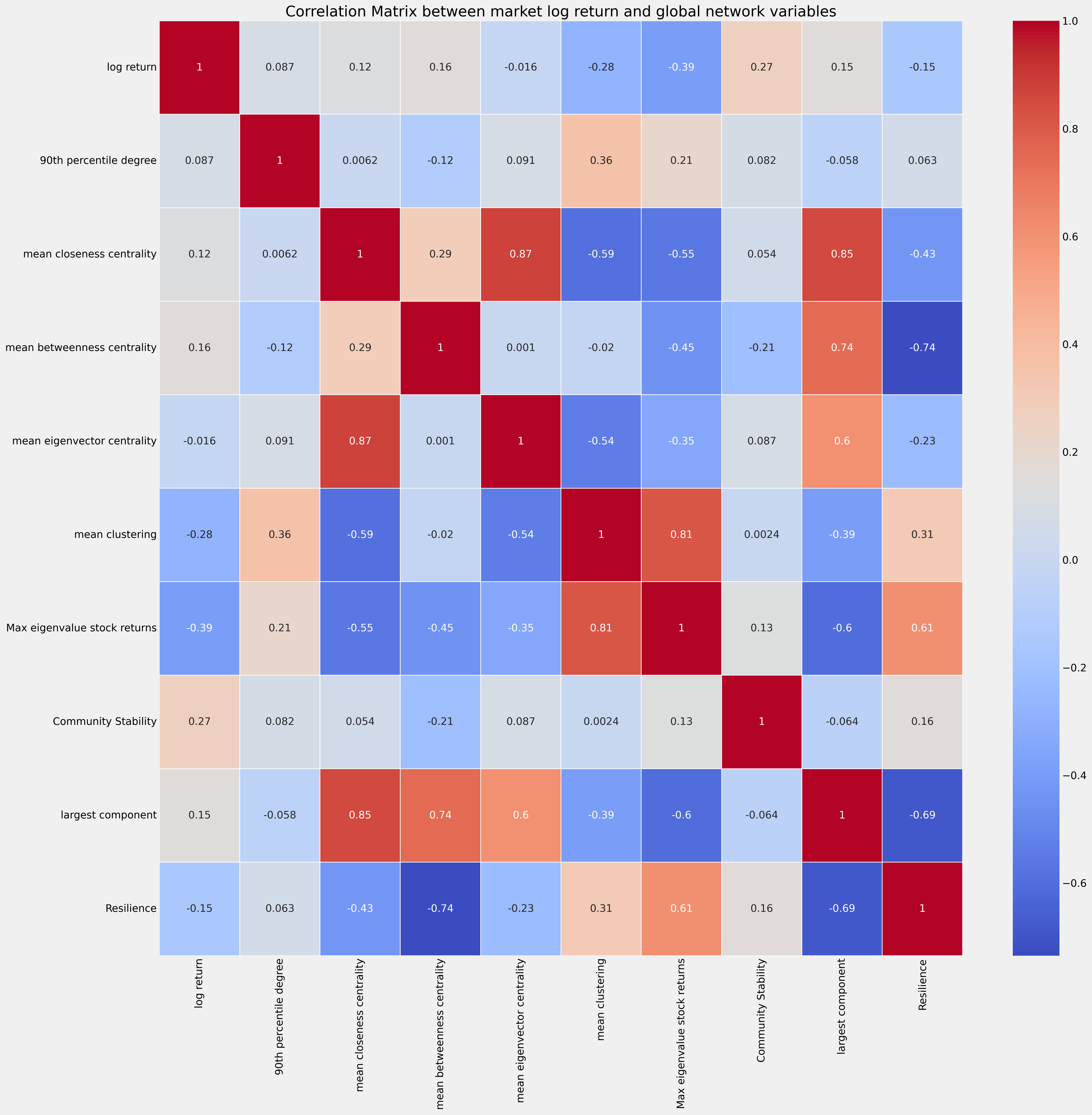}
    \caption{Correlation matrix of the log-return of all S$\&$P stocks with the network properties on 30 observations (in years)}
    \label{fig:corrglobal}
\end{figure}

In Fig.\ref{fig:corrglobal} we compute the correlation matrix of the global network properties and the market log-return using the 30 time steps (years). We observe that some variables are anti-correlated to the log return with an absolute value greater than $20\%$: the mean clustering and the max eigenvalue of the stock returns. This suggests that when stocks are highly connected to one another (mean clustering) or when they follow the market trend (largest eigenvalue), the global log return decreases.  On the contrary, the log return is positively correlated with a coefficient greater than $20\%$ with the community stability variable.

\begin{table}
    \centering
    
    \begin{tabular}{lcc}
        \toprule
        \textbf{Variable} & \textbf{Lag} & \textbf{SSR F-test p-value} \\
        \midrule
        Mean Closeness Centrality & 3 & 0.0424 \\
                                 & 4 & 0.0352 \\
        \midrule
        \addlinespace
        Community Stability & 1 & 0.0043 \\
                            & 2 & 0.0244 \\
        \bottomrule
    \end{tabular}
    \caption{Summary of significant SSR F-test $p$-values ($p < 0.05$) for Granger causality on log return with the global network variables on the long term period.}
\end{table}\label{tab:granger1}

Given the limited number of observations (Nobs=30) it is challenging to discuss the correlations between network evolution and stock returns. Nonetheless we perform a Granger causality test \cite{granger1969investigating} using a maximum lag of 5 years as an indicator to test if the time series of the network properties can use to predict the stock returns. We only quote the result for the most restrictive statistical test which corresponds to the Sum of Squared Residuals (SSR) F-test for computing the p-value\footnote{other statistical tests for the SSR are $\chi ^2$-test and Likelihood ratio. While we obtain $p$-values less than 0.05 in some cases, we choose to disregard them.}. We obtain a $p$-value less than 0.05 for the measurement given in Tab.\ref{tab:granger1}, suggesting that these variable could be used as predictors for forecasting the global stock returns, especially the community stability at lag 1.

\subsection{The short term period}

\begin{figure}
    \centering
    \includegraphics[width=1\linewidth]{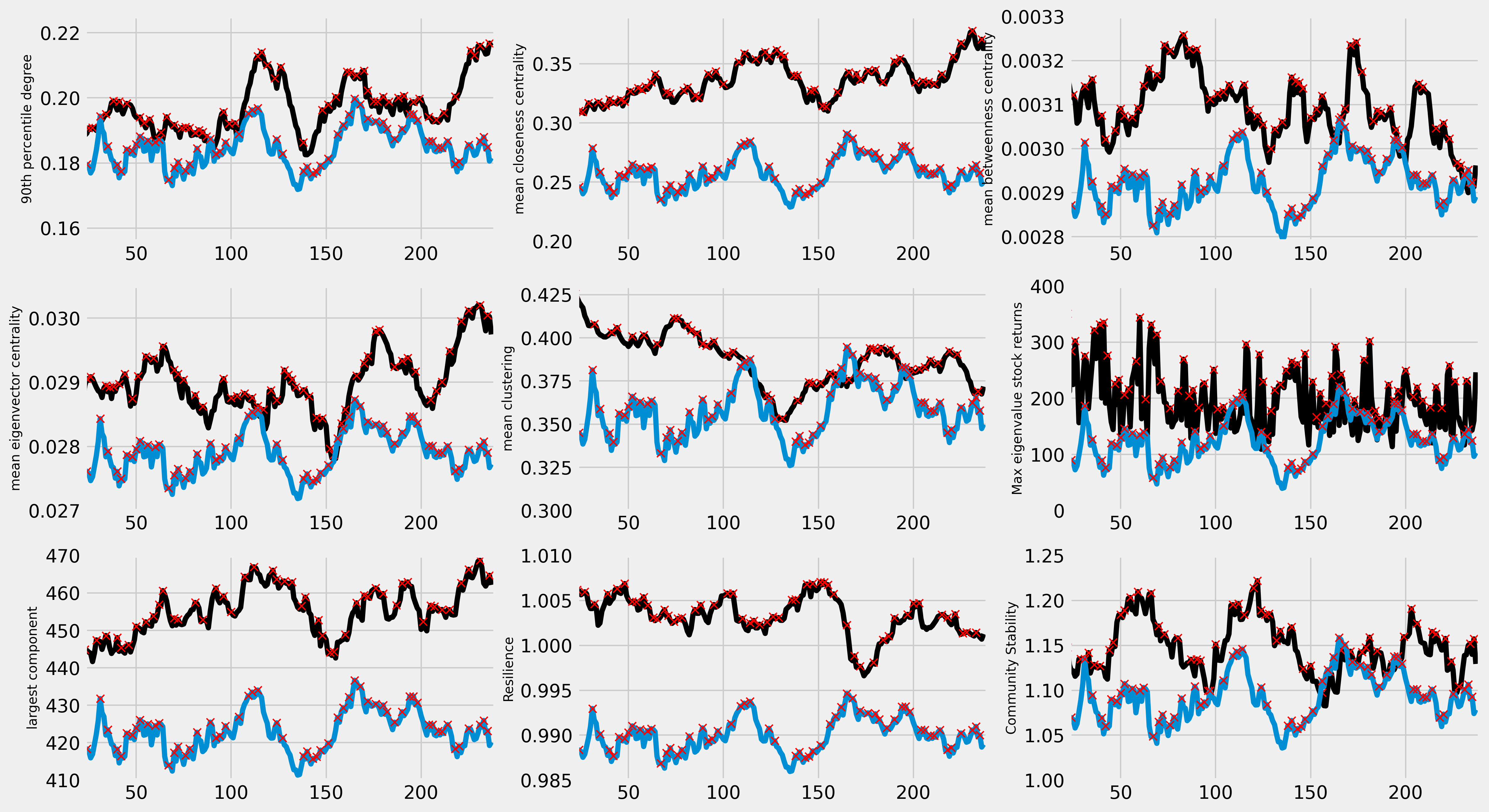}
    \caption{Evolution of the network properties every two business days smoothed over a rolling window $W=23$. The blue curves correspond to the stock returns rescaled. The number of observed time scale is $Nobs$=238. The red crosses correspond to peaks in the curves.}
    \label{fig:globaldyn_h}
\end{figure}

\begin{figure}
    \centering
    \includegraphics[width=0.99\linewidth]{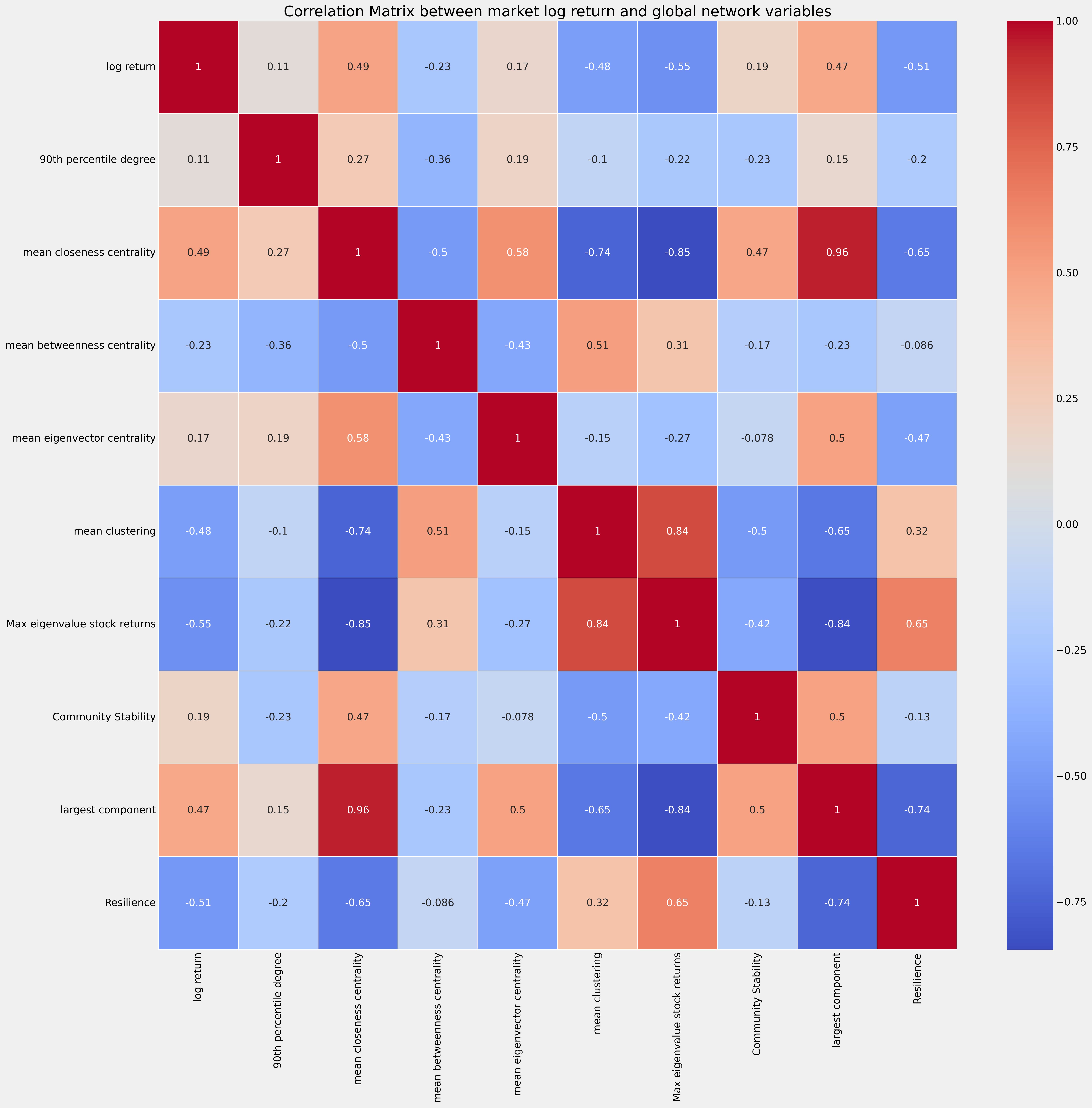}
    \caption{Correlation matrix of the log-return of all S$\&$P stocks with the network properties on 238 observations (every two days)}
    \label{fig:corrglobal_hours}
\end{figure}

In Fig.\ref{fig:globaldyn_h} we display the global dynamical evolution of the network properties (black curves) and the rescaled global stock returns smoothed with a rolling mean average window $W=23$. The matrix correlation is displayed in Fig.\ref{fig:corrglobal_hours}. 

In the short time scale,  the top 5 most correlated variables (in absolute value) are:
\begin{itemize}
    \item The max eigenvalues of the stock returns, with a coefficient of -0.55, also the highest correlation on long term period. 
    \item The mean closeness centrality, with a coefficient of 0.49
    \item The mean clustering, with a coefficient of -0.48, which was the second highest coefficient for the long term period. 
    \item The largest component, with a coefficient of 0.47
    \item The resilience, with a coefficient of -0.35. 
\end{itemize}

This is quite interesting, pointing out that the dynamical processes between the stocks on very different time scales can be partially captured by the maximum eigenvalues of the stock returns and the mean clustering of the network. The former is not surprising as it captures the market mode \cite{laloux1999noise} and it was previously shown that high eigenvalues are an indicator of financial crisis \cite{plerou2000random}. The latter was also studied in \cite{boginski2005statistical}, \cite{marti2017review}, \cite{onnela2003dynamics} as a metric that can reflect the underlying stability of the market. Here we find that they are consistent over very different time scales when studying the global stock returns.

\begin{table}[h]
    \centering
    \begin{tabular}{lcc}
        \toprule
        \textbf{Variable} & \textbf{Best Lag} & \textbf{SSR F-test p-value} \\
        \midrule
        90th Percentile Degree & 2 & $2.16 \times 10^{-10}$ \\
        \midrule
        Mean Closeness Centrality & 2 & 0.00939 \\
        \midrule
        Mean Betweenness Centrality & 3 & $8.67 \times 10^{-8}$ \\
        \midrule
        Mean Eigenvector Centrality & 6 & 0.00370 \\
        \midrule
        Mean Clustering & 3 & 0.00039 \\
        \midrule
        Max Eigenvalue Stock Returns & 1 & $4.09 \times 10^{-8}$ \\
        \midrule
        Community Stability & 1 & $4.50 \times 10^{-5}$ \\
        \midrule
        Largest Component & 2 & $5.00 \times 10^{-7}$ \\
        \midrule
        Resilience & 3 & $1.51 \times 10^{-11}$ \\
        \bottomrule
    \end{tabular}
    \caption{Best Lag and SSR F-test p-value for each global network Variable on the short time period.}\label{tab:grangerglobh}
\end{table}

The Granger causality test \cite{granger1969investigating} result for the best lag of our network variables is reported in  Tab.\ref{tab:grangerglobh}, suggesting  again that these variables could be used as predictors for forecasting the global stock returns. Again, it is interesting to compare these variables with the ones we find on the long-term period. For the long term period we only find the mean closeness centrality and the community stability as meaningful properties to predict the stock returns. In the short time period, we find that all 9 variables (with different lag) are correlated with the future stock return. The best $p$-value being for the Resilience at lag 3 and the 90th percentile degree at lag 2. Here the number of observation is an order of magnitude larger which makes this test more robust.

\section{Evolution of the individual properties of stocks within the network: application to forecasting stock returns}\label{secindiv}

Previously we have studied how the market log return correlates in time with respect to global properties of the network. We find some interesting correlations with these properties suggesting some predictability and a deeper understanding of how stocks interact with each other, leading to a coherent collective behaviour. Now we might wonder how these interactions impact individual stocks: by considering the position and properties of stock $i$ within the network, along with the global properties of the network, can we predict its log return more accurately than by using standard variables? 

To address this question, we perform a forecast analysis using the following individual variables for each stock: 1- its degree centrality, 2-its closeness centrality, 3-its betweenness centrality, 4-its eigenvector centrality and 5-its clustering coefficient. These variables characterize the stock within the network. In addition we take the global network variables that we previously consider. They contain information about the overall market structure.

The standard variable that we consider for the baseline scenario is the stock's log return at the previous time step (lag).

\subsection{Methodology}\label{method}
\subsubsection{Preparing the data.}
We start by randomly selecting 85$\%$ of the stocks for training and $15\%$ for testing. For each stock we smooth all variable $f(t)$ (including the log return) using a rolling mean average of window $W$, defined as: 
\begin{equation}
    \bar{f}(t)=\frac{1}{W} \sum_{i=0}^{W-1} f(t-i)
\end{equation}
where $\bar{f}(t)$ is the rolling mean of variable $f$ at time $t$ and $f(t-i)$ is the value of variable $f$ at time $t-i$. Note that one could adapt the average window to each variable by performing a grid search on the best forecasting results for instance. In what follows we adopt a fixed criterion for the value $W$, taking 10$\%$ of the number of the observed time. 

We further compute all variables at different lag with a maximum given by Nlag. We then remove the time scale $t<Nlag$. Finally we concatenate all training stocks into one training dataset, implicitly assuming that the time series are approximately stationary. 
\subsubsection{Selecting the predictive variables.}
For each variable in the training dataset we compute its correlation with the target variable (the log return) and store its values. We then select a variable if its correlation with the log return is greater than the 65th percentile of the correlation values.  

\subsubsection{The regression models.}
We consider 5 models: 1- a Gradient Boosting Regressor model \cite{friedman2001greedy} using all selected variables (GBR), 2- a Random Forest Regressor \cite{breiman2001random} using all selected variables (RFR), 3- a linear regression model using solely the lag 1 of the log return as input variable (LRbase), 4- a Random Forest Regressor using as input  variables of the selected lag of the log return (RFRbase), 5- a weighted average of the first 2 models (wA) where the weight is given by the r2 score.    

\subsubsection{Predictions and scoring.}
We use the 15$\%$ of stocks (Ntest=Nstocks*0.15) not used for training and predict their log return on the entire time scale: $Nobs-Nlag$. For each stock prediction we compute the F2score \cite{powers2011evaluation} and the mean absolute error \cite{willmott2005advantages} to build their distribution for each regression model.

\subsection{Long time scale forecasting}

We start with the selected S$\&$P500 stocks from 1993-01-01 to 2024-01-01 (267 stocks) where we have Nobs=30 (one measurement for each business year) with Nlag=5 such that the training sample Ntrain=$(30-5)\times 85\% \times 267$=5673. We use a window $W=10\%Nobs=3$ to compute the rolling mean. 

We perform a Granger causality test on the network variables and report the best lag $p$-value for the SSR F-test in Tab.\ref{tab:granger2}.
\begin{table}
\centering
\begin{tabular}{lcc}
\toprule
\textbf{Variable} & \textbf{Best Lag} & \textbf{SSR F-test p-value} \\
\midrule
Degree Centrality & 1 & 0.0033\\
\midrule
Closeness Centrality & 1 & $1.86 \times 10^{-15}$ \\
\midrule
Betweenness Centrality & 1 & $6.55 \times 10^{-5}$ \\
\midrule
Eigenvector Centrality & 1 & 0.02 \\
\midrule
Clustering & 3 & $5.96 \times 10^{-10}$ \\
\midrule
90th percentile degree & 1 & $1.57 \times 10^{-23}$ \\
\midrule
mean closeness centrality & 1 & $9.28 \times 10^{-40}$ \\
\midrule
mean betweenness centrality & 4 & $7.55 \times 10^{-63}$ \\
\midrule
mean eigenvector centrality & 4 & $1.1 \times 10^{-28}$ \\
\midrule
mean clustering & 3 & $6.67 \times 10^{-28}$ \\
\midrule
Max eigenvalue & 2 & $3.35 \times 10^{-25}$ \\
\midrule
Community Stability & 3 & $3.35 \times 10^{-27}$ \\
\midrule
largest component & 4 & $2.54 \times 10^{-71}$ \\
\midrule
Resilience & 4 & $1.65 \times 10^{-54}$ \\
\bottomrule
\end{tabular}
\caption{Best lag and SSR F-test $p$-value for each variable for the long time scale forecasting}\label{tab:granger2}
\end{table}
The first 5 variables are specific to each stocks while the others are the global variables of the network and the max eigenvalue of the stock return correlation matrix. The result is interesting as it confirms that a stock's network properties do correlate with its future value at lag 1 but also the global properties of the network at higher lags with lower p-values.

The selected variables for the training are reported in the appendix Tab.\ref{tab:selecvariablelong}. The first two most correlated variables are the log return at lag 1 and 2 followed by the 90th percentile degree at lag2. The most correlated network variable that is not a global network variable is the closeness centrality at lag3. In fact over the selected network variables, 6 of them are characteristic to individual stocks and 17 describe the global network properties.

\begin{figure}
    \centering
    \includegraphics[width=0.9\linewidth]{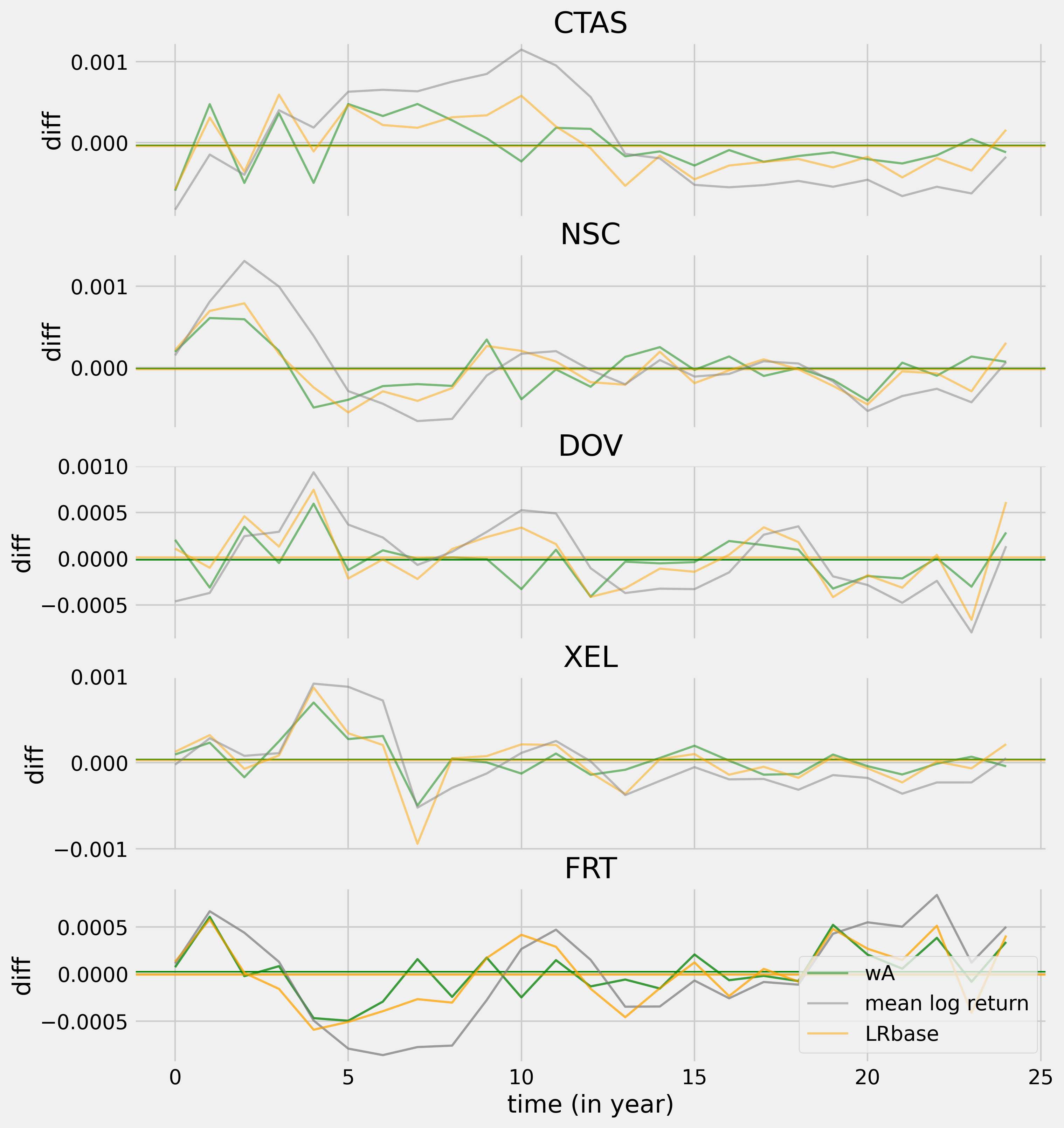}
    \caption{Differences between the wA prediction and the log return (green curves), the LRbase and the log return (orange curves) and the mean of the log return with itself (grey curves) for 5 randomly selected stocks from the testing set. The horizontal lines correspond to the mean of the differences.}
    \label{fig:sample_difflong}
\end{figure}

In Fig.\ref{fig:sample_difflong} we show the differences between the wA prediction and the log return (green curves), the LRbase and the log return (orange curves) and the mean of the log return with itself (grey curves) for 5 randomly selected stocks from the testing set. The horizontal lines correspond to the mean of the differences. Qualitatively, the differences between LRbase and wA are not significant but we observe that the two models do better than a simple mean average of the log return.% which is why the median R2 score is not null but positive. 

\begin{figure}
    \centering
    \includegraphics[width=0.9\linewidth]{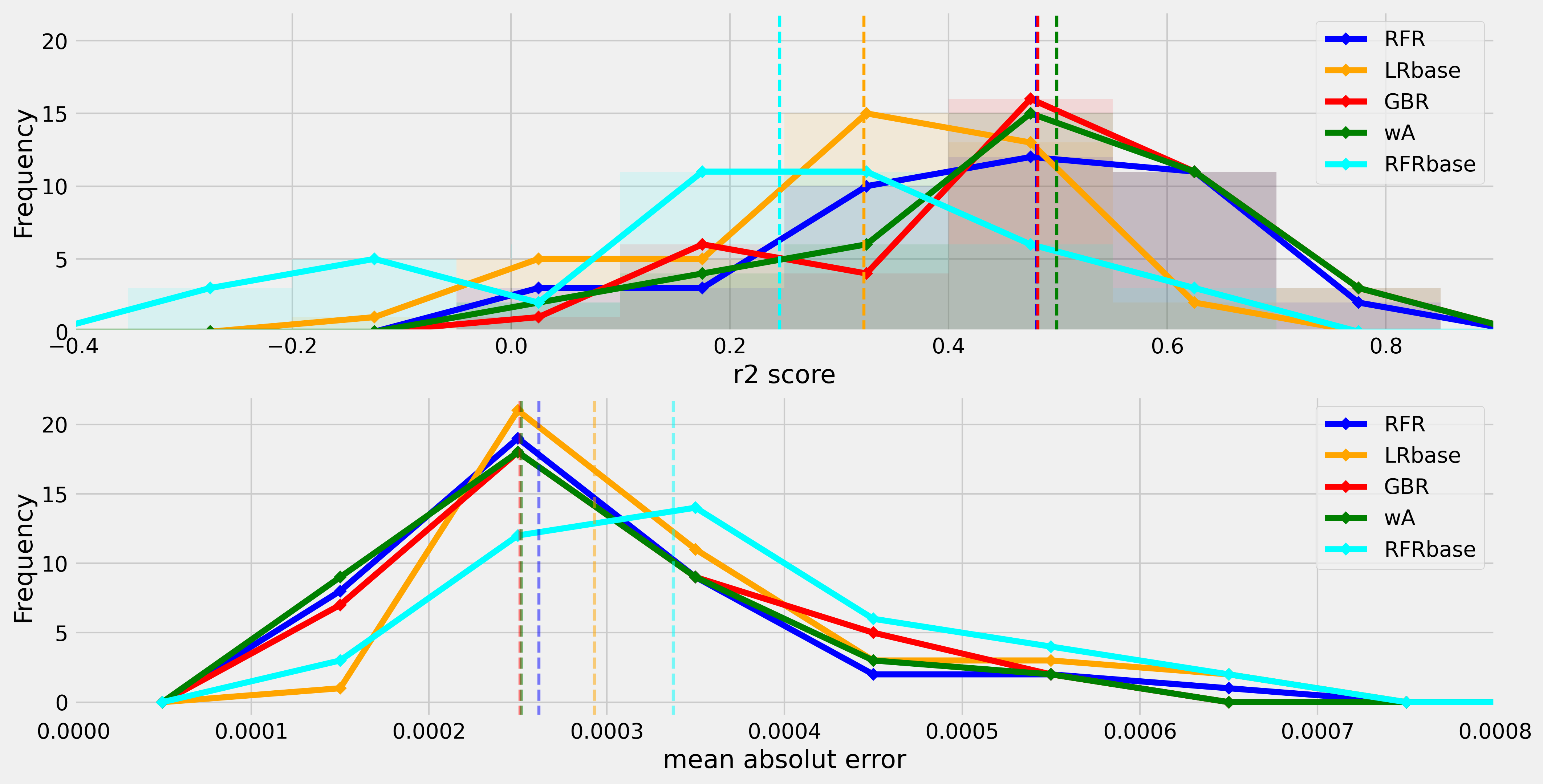}
    \caption{R2 distribution (top panel) and MAE distribution (lower panel) computed using the testing set of stocks for the long time scale forecasting. Vertical dashed lines correspond to the median values.}
    \label{fig:R2MAElong}
\end{figure}
For each stock of the testing set we measure the R2 score and the MAE to compute their distribution for our different models that we display in Fig.\ref{fig:R2MAElong}. Firstly we observe the wide skewed distribution of the R2 score and MAE point out the high variability of these simple predictive models. Nonetheless the question we want to address is if adding the network input variables improve the overall prediction of the individual stock returns. The vertical dashed lines correspond to the median of the distribution. The values are [0.48,0.32,0.48,0.5,0.25] for the [RFR, LRbase,GBR,wA,RFRbase] respectively. If we wish to compare the added value of the network parameters we shall compare the improvement of the RFRbase (which contains only the selected lag of the log return) with the RFR which contains only the selected lag of the log return and the network parameters at the selected lag). However, it seems like the LRbase (which contains only lag 1 of the log return) provides a better fit than the RFRbase. In this case we have an improvement of $0.48/0.32-1=50\%$ on the R2 score ($56\%$ improvement if we consider the wA model). For the MAE the values are [0.00026,0.00029,0.00025,0.00025,0.00034] for [RFR, LRbase,GBR,wA,RFRbase] respectively.

\subsection{Short time scale forecasting}
Similarly to the long-term forecast we use the rolling mean average window of size W=int($10\%Nobs)=23$ and for the number of lag we take $Nlag=12$ after checking that including higher lag does not change the accuracy of the prediction.

Following the same methodology as described in \ref{method}, we report in Tab.\ref{tab:grangerlocalh} the Granger causality test results on the best Lag. 

\begin{table}[h]
    \centering
    \begin{tabular}{lcc}
        \toprule
        \textbf{Variable} & \textbf{Best Lag} & \textbf{SSR F-test p-value} \\
        \midrule
        Degree Centrality & 9 & $2.43 \times 10^{-13}$ \\
        \midrule
        Closeness Centrality & 11 & $6.40 \times 10^{-37}$ \\
        \midrule
        Eigenvector Centrality & 10 & $5.12 \times 10^{-13}$ \\
        \midrule
        Clustering & 9 & $2.49 \times 10^{-22}$ \\
        \midrule
        90th percentile degree & 10 & 0\\
        \midrule
        mean closeness centrality & 11 & $2.23 \times 10^{-111}$ \\
        \midrule
        mean betweenness centrality & 10 & $3.22 \times 10^{-94}$ \\
        \midrule 
        mean eigenvector centrality & 10 & $3.31 \times 10^{-151}$ \\
        \midrule 
        mean clustering & 11 & $1.85 \times 10^{-154}$ \\
        \midrule 
        Max eigenvalue & 11 &  $8.29 \times 10^{-186}$ \\
        \midrule 
        Community Stability & 11 &  $1.53 \times 10^{-232}$ \\
        \midrule  
        largest component & 8 &  $6.86 \times 10^{-127}$ \\
        \midrule  
        Resilience & 11 &  $6.54 \times 10^{-194}$ \\
        \bottomrule
    \end{tabular}
    \caption{Best lag and SSR F-test $p$-value for each variable for the short time scale forecasting.}\label{tab:grangerlocalh}
\end{table}

 In the this case, the first 4 variables are associated with the individual properties of the stock within the network while the others are global properties of the network and the max eigenvalue of the stock return correlation matrix. Interestingly, the global variables have a lower p-value than the individual ones, suggesting that the market structure has more weight in the forecast of the individual stocks compared to the interactions of the stock within the network.

The variables most correlated with the log return that are used for the training are reported in the appendix Tab.\ref{tab:selecvariableshort}. In this case the most correlated variables are the log return itself at lag=[1,2,..12] with a correlation coefficient ranging from [0.95,0.9,...,0.44] respectively. None of the variables selected for the training are individual network variable expect for the Closeness Centrality at lag 2 and 3 that appears at rank 48 and 59 with a correlation coefficients equal to 0.09 and 0.08 suggesting again that the global network properties have more impact on the individual stock return than the stock properties within the network. Interestingly, the closeness centrality was also the most correlated individual variable in the long time period.

In the appendix Fig.\ref{fig:difflrhours} we also show the differences of the predicted log return for the wA and LRbase for 5 randomly selected stocks from the testing set.

\begin{figure}
    \centering
    \includegraphics[width=0.9\linewidth]{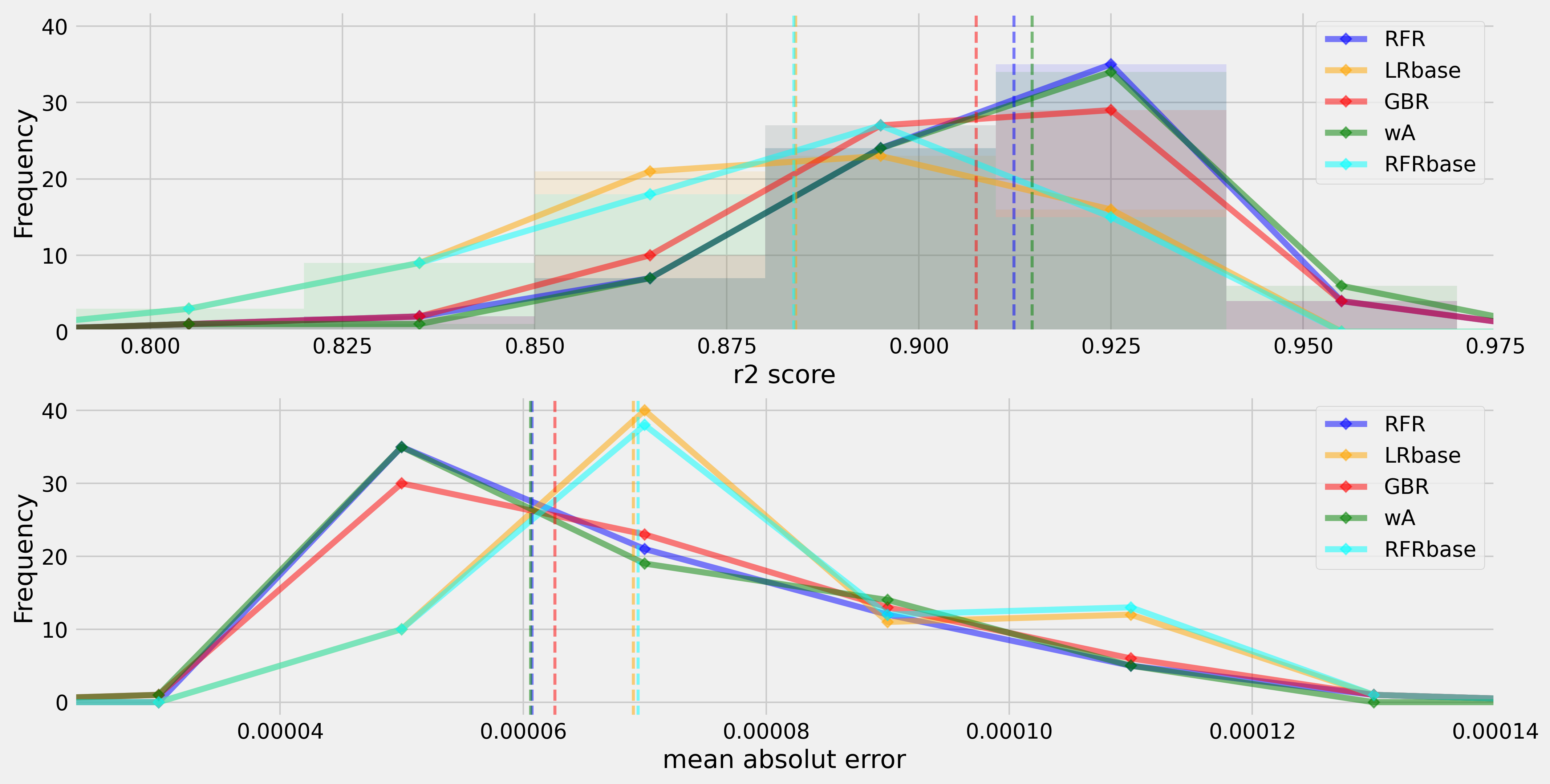}
    \caption{R2 distribution (top panel) and MAE distribution (lower panel) computed using the testing set of stocks for the long time scale forecasting. Vertical dashed lines correspond to the median values.}
    \label{fig:R2MAEshort}
\end{figure}

The resulting forecast agreement on the testing set is displayed in Fig.\ref{fig:R2MAEshort}. For these predictions the distribution of the R2 scores and the MAE are much better than for the long time scale. The training sample size and the higher rolling window are probably the main reasons why. In any case, we report the median score for the R2 distributions are [0.91,0.88,0.91,0.91,0.88] for the [RFR, LRbase,GBR,wA,RFRbase] respectively. Hence the improvement brought by the network variables is 0.91/0.88-1=3$\%$. For the MAE median scores we have [$6\times 10^{-5},7\times 10^{-5},6\times 10^{-5},6\times 10^{-5},7\times 10^{-5} $] or the [RFR, LRbase,GBR,wA,RFRbase] respectively. \footnote{Note that we also test the forecasting using a rolling average window of size W=10, for which the median R2score are [0.84,0.78,0.82,0.84,0.79], leading to a $6\%$ improvement, for W=30: [0.93,0.91,0.93,0.93,0.91] leading to a $2\%$ improvement and for W=5: [0.72,0.59,0.7,0.73,0.6] leading to a 21$\%$ improvement.}

\subsection{Discussion}
We find some interesting results in both the long and the short time periods. First, as expected the log return itself at previous lag is most important variable to predict the future value of a stock. Second, the global network variable are more correlated to the future return of a stock than the stock characteristics within the network. This suggests that the overall market structure plays a more important role than the stock's interactions within the network. The closeness centrality measure of a stock is the most important individual feature to predict the future return of a stock in our models. The resilience of the network and the 90th Percentile Degree are also strong global features that correlate with the future return of a stock. 
Overall the network variables improve the forecasting of individual stock returns both on short and long time period.

\section{Summary and conclusions}\label{secconclu}

In this article we study the dynamical evolution of a network made from stock returns correlations and how its properties correlate with the log return itself. We consider global properties of the network as well as individual properties of the nodes. This allows us to study the correlation between 1- the market return evolution with the global properties of the network, and 2- the evolution of individual stock returns with the characteristics of the stock within the network. Both at the collective level and on the individual level we find meaningful indicators for the future evolution of the stocks (Granger causality test). We also study the dynamical evolution on two different time scales: long period and short period. We find that some indicators are scale invariant (the largest eigenvalue of the correlation matrix and the average clustering of the network) for the collective evolution. For the individual stocks, the degree of the node, the closeness centrality, the eigenvector centrality and the clustering of the node are meaningful indicators on both long- and short-term periods, although they were not selected to forecast the future value of our testing stocks for the short time period except for the closeness centrality measure.  

Finally, we use the network variables to forecast the evolution of stock returns and find an improvement over baseline scenarios that do not include network features (50$\%$ improvement for the long period and $3\%$ for the short period). This indicates that network variables capture some of the interactions of the complex financial system, thus providing an added value towards predicting the dynamical evolution of collective and individual stock returns. 

This work could be expanded in several ways. For instance one could add additional variables in the training dataset when predicting individual stock returns, such as the variance of some of the network variables. It would also be interesting to extend the time period of the data and test some more sophisticated forecasting models.

\section*{Data and code declaration}
The data used for this project can be downloaded from $Yahoo$ finance's API. The code used to generate this analysis can be provided upon request and will be posted online after revision of this manuscript at \url{https://github.com/IxandraAchitouv}

\section*{Appendix: additional figures and tables}

\begin{figure}
    \centering
    \begin{minipage}[b]{0.45\textwidth}
        \centering
        \includegraphics[width=\textwidth]{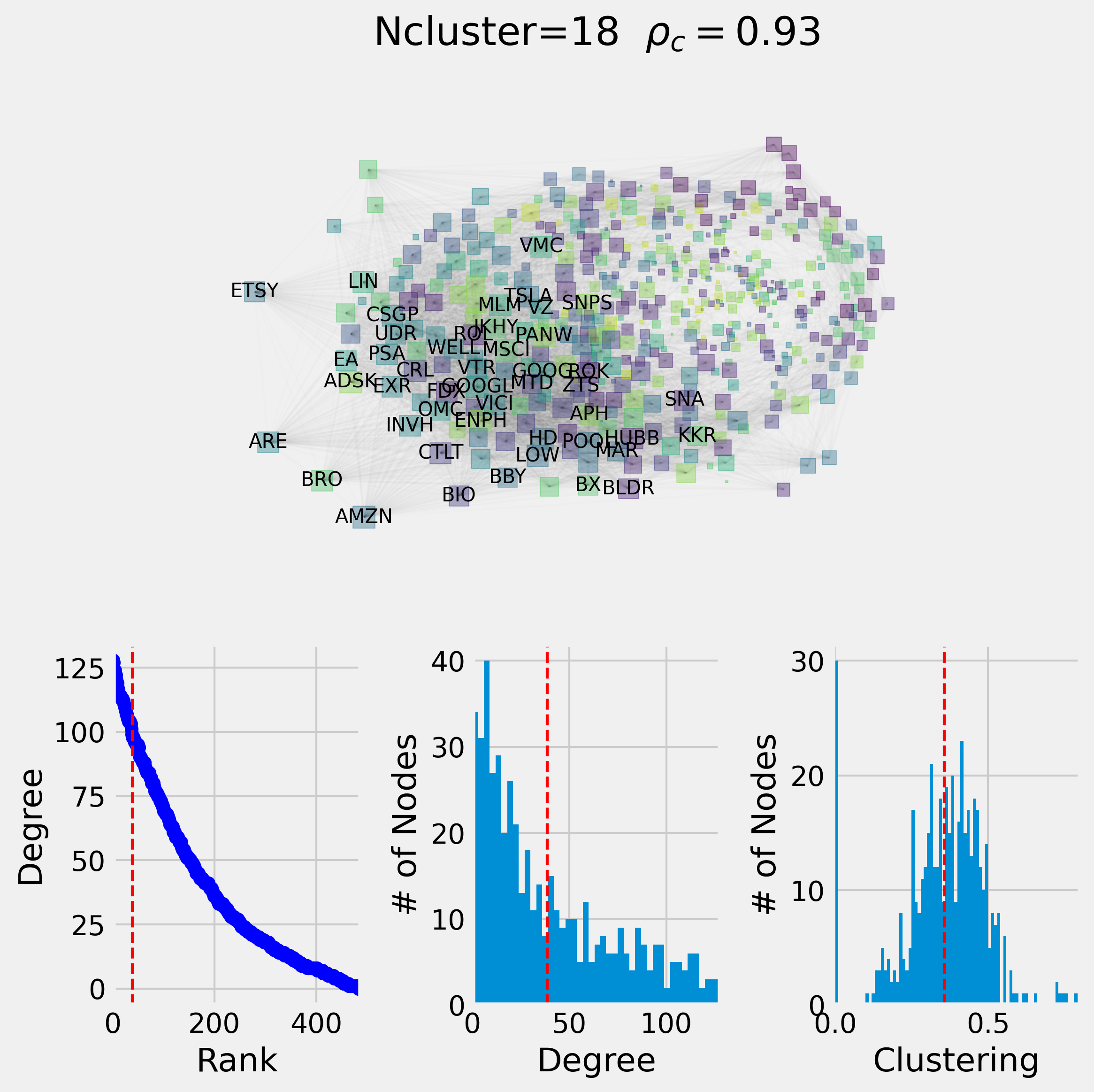}
        %\caption{Caption for Figure 1}
        \label{fig:1}
    \end{minipage}
    \hfill
    \begin{minipage}[b]{0.45\textwidth}
        \centering
        \includegraphics[width=\textwidth]{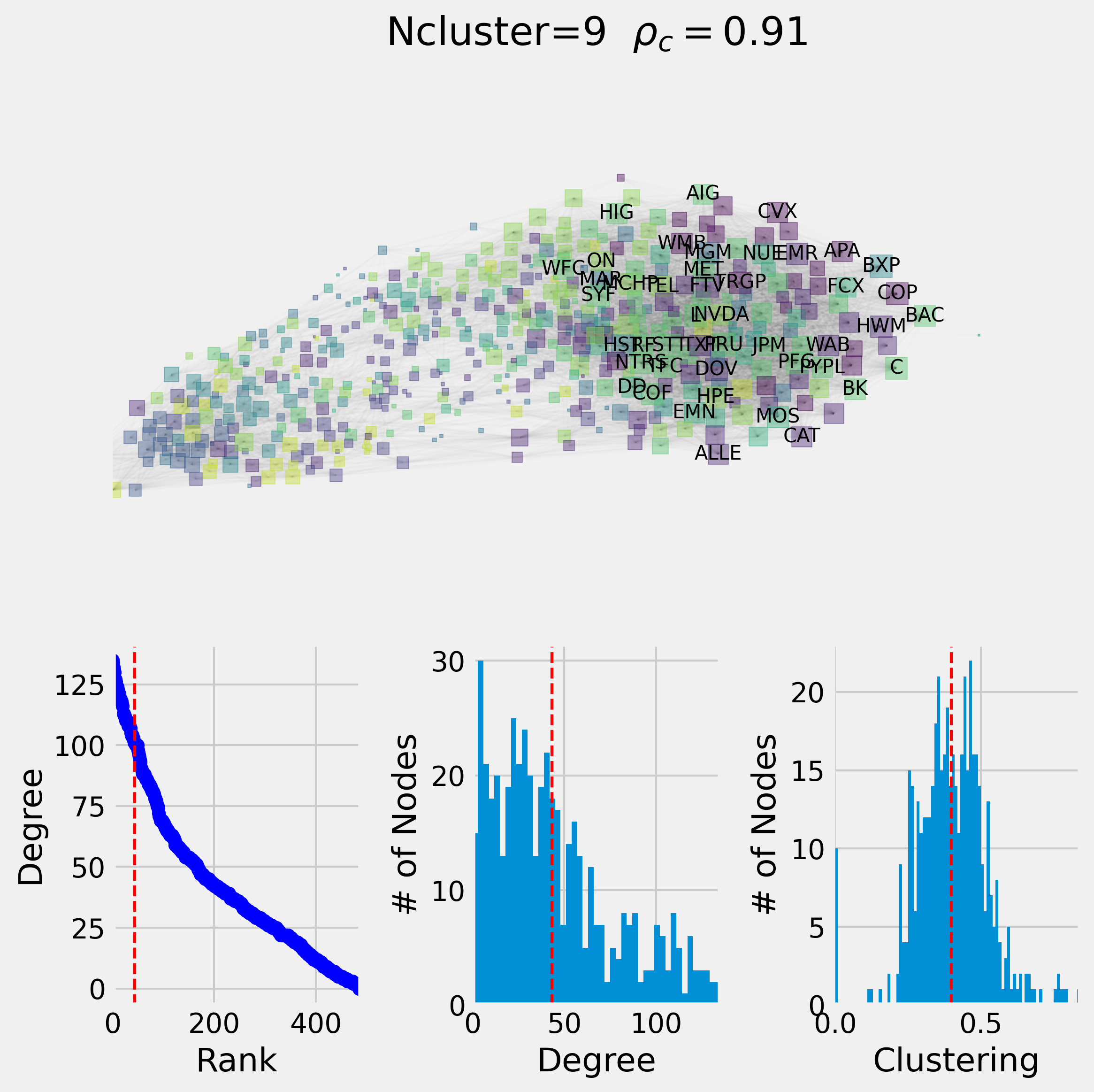}
        %\caption{Caption for Figure 2}
        \label{fig:2}
    \end{minipage}
    \vfill
    \begin{minipage}[b]{0.45\textwidth}
        \centering
        \includegraphics[width=\textwidth]{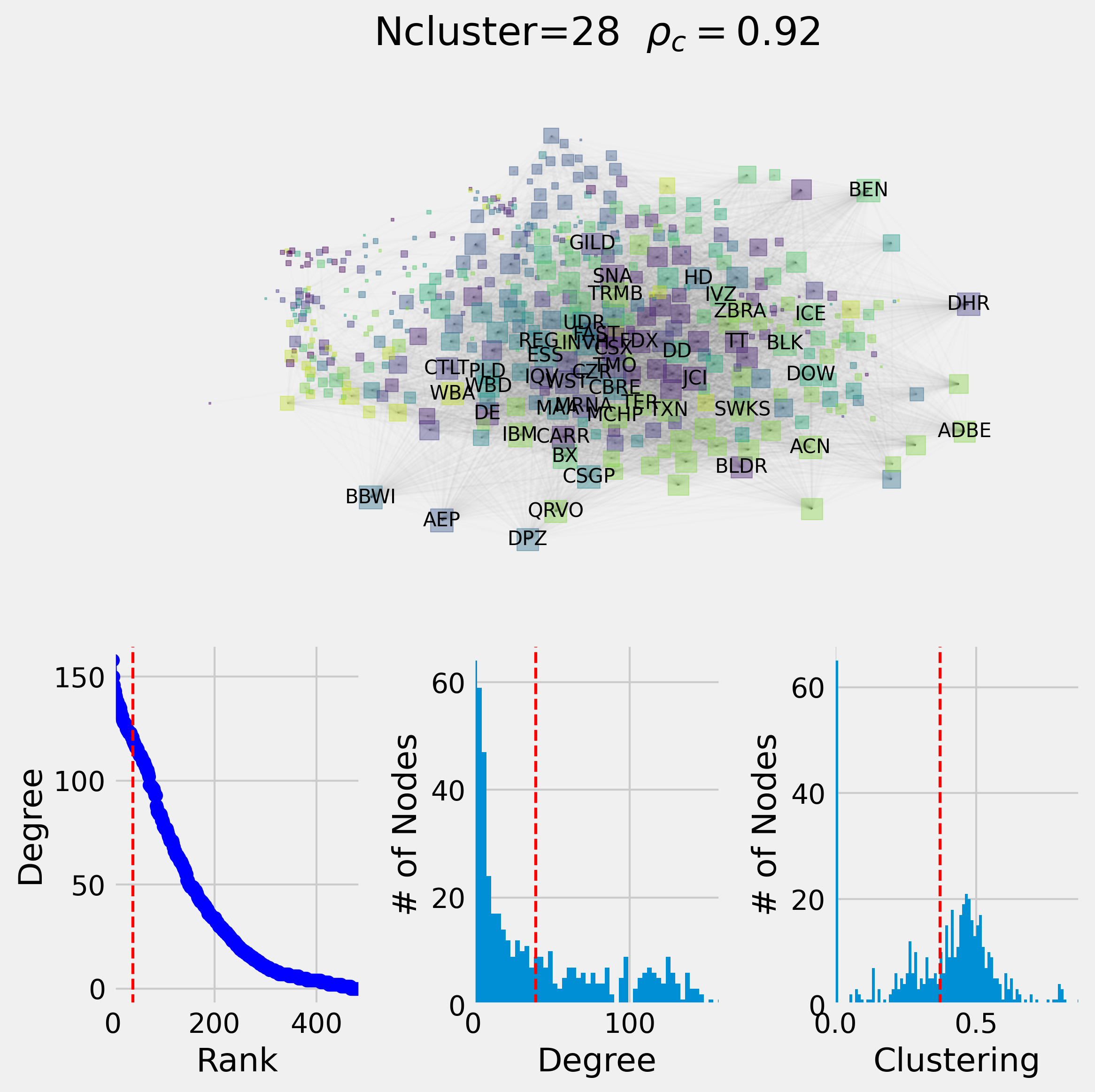}
       % \caption{Caption for Figure 3}
        \label{fig:3}
    \end{minipage}
    \hfill
    \begin{minipage}[b]{0.45\textwidth}
        \centering
        \includegraphics[width=\textwidth]{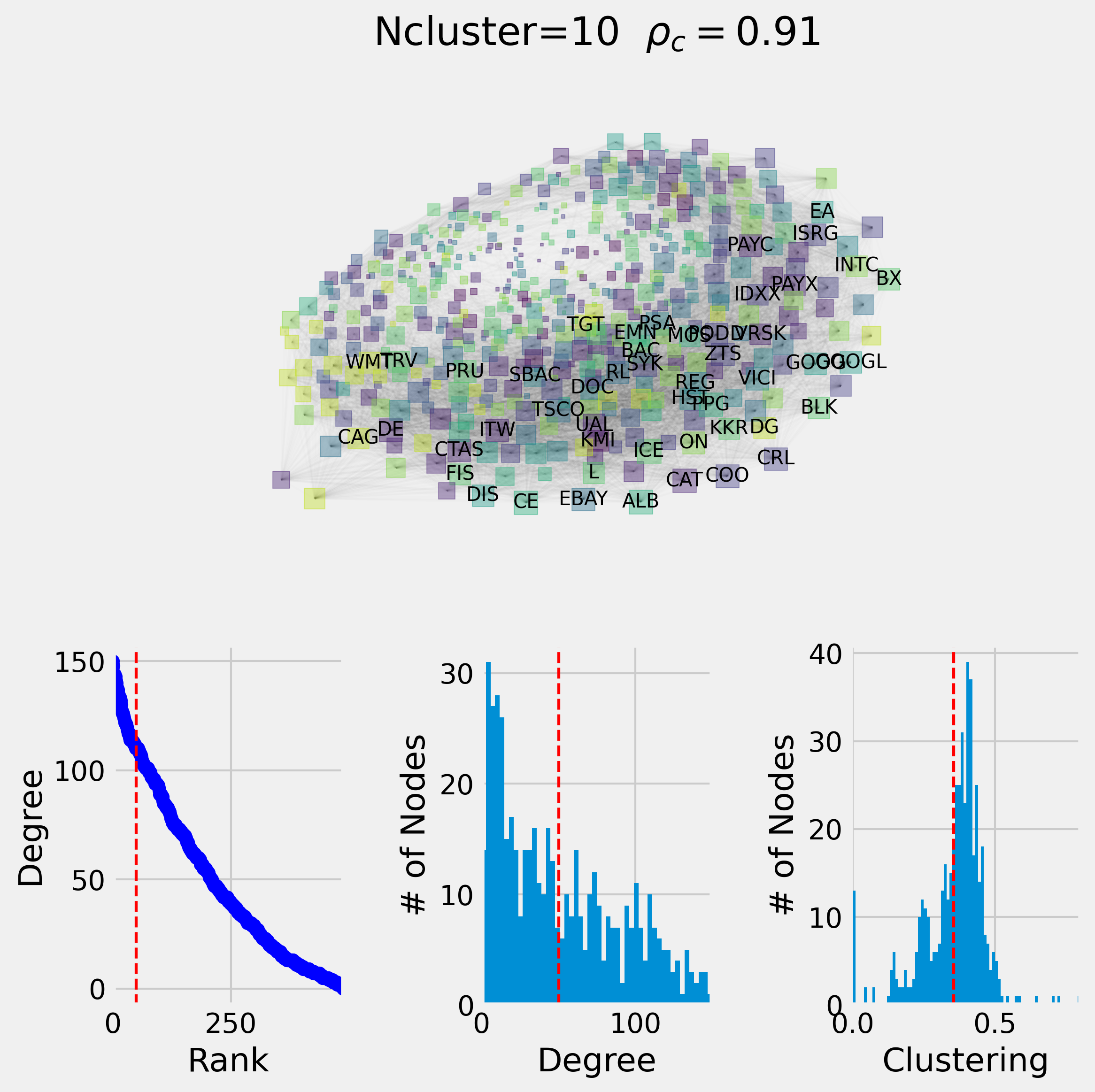}
       % \caption{Caption for Figure 4}
        \label{fig:4}
    \end{minipage}
    \caption{Network of the stocks for different short time scales: [46,106,151,211] (top panels) along with the degree distribution (left bottom panels), eigenvector centrality histogram (middle bottom panels) and clustering histogram of the stocks (right bottom panels). Red vertical lines correspond to the mean.}
    \label{fig:network_sampleh}
\end{figure}
In Fig.\ref{fig:network_sampleh} we display the network built on the short time scale for 4 randomly selected scale. Compared to Fig.\ref{fig:network_sample} the number of stocks are higher. We observe some clustering per sector (shown by the colors). In average the number of clusters found with the Louvain is smaller than in the long time period. 

\begin{table}[h]
\centering
\small
\begin{tabular}{|c|l|c|} % Adjust column width
\hline
\textbf{Rank} & \textbf{Variables} & \textbf{Correlation coefficients} \\ \hline
26 & Mean Closeness Centrality\_3 & 0.09 \\ \hline
25 & Clustering\_5 & 0.09 \\ \hline
24 & Max Eigenvalue Stock Returns\_3 & 0.09 \\ \hline
23 & Mean Clustering\_2 & 0.09 \\ \hline
22 & Resilience\_5 & 0.09 \\ \hline
21 & Eigenvector Centrality\_5 & 0.09 \\ \hline
20 & Resilience\_2 & 0.09 \\ \hline
19 & Mean Eigenvector Centrality\_4 & 0.09 \\ \hline
18 & Max Eigenvalue Stock Returns\_2 & 0.10 \\ \hline
17 & Degree Centrality\_4 & 0.10 \\ \hline
16 & 90th Percentile Degree\_4 & 0.10 \\ \hline
15 & 90th Percentile Degree\_5 & 0.10 \\ \hline
14 & Closeness Centrality\_2 & 0.11 \\ \hline
13 & Degree Centrality\_5 & 0.11 \\ \hline
12 & Resilience\_1 & 0.11 \\ \hline
11 & Mean Eigenvector Centrality\_5 & 0.11 \\ \hline
10 & Closeness Centrality\_3 & 0.12 \\ \hline
9  & Resilience\_4 & 0.12 \\ \hline
8  & Mean Closeness Centrality\_5 & 0.13 \\ \hline
7  & Largest Component\_5 & 0.13 \\ \hline
6  & Largest Component\_2 & 0.15 \\ \hline
5  & 90th Percentile Degree\_2 & 0.15 \\ \hline
4  & Mean Closeness Centrality\_2 & 0.17 \\ \hline
3  & 90th Percentile Degree\_1 & 0.22 \\ \hline
2  & Log Return\_2 & 0.28 \\ \hline
1  & Log Return\_1 & 0.63 \\ \hline
\end{tabular}
\caption{Selected variables for the training of the long time period}
\label{tab:selecvariablelong}
\end{table}

\begin{table}[h]
\centering
\begin{tabular}{|c|l|c|}
\hline
\textbf{Rank} & \textbf{Variables} & \textbf{Correlation coefficients} \\ \hline
42 & Resilience\_9 & 0.10 \\ \hline
41 & Largest Component\_3 & 0.10 \\ \hline
40 & Max Eigenvalue Stock Returns\_11 & 0.10 \\ \hline
39 & Resilience\_8 & 0.11 \\ \hline
38 & Mean Closeness Centrality\_3 & 0.11 \\ \hline
37 & 90th Percentile Degree\_4 & 0.11 \\ \hline
36 & Max Eigenvalue Stock Returns\_10 & 0.11 \\ \hline
35 & Resilience\_7 & 0.11 \\ \hline
34 & Max Eigenvalue Stock Returns\_9 & 0.11 \\ \hline
33 & Max Eigenvalue Stock Returns\_8 & 0.12 \\ \hline
32 & Largest Component\_2 & 0.12 \\ \hline
31 & 90th Percentile Degree\_3 & 0.12 \\ \hline
30 & Mean Closeness Centrality\_2 & 0.12 \\ \hline
29 & Resilience\_6 & 0.12 \\ \hline
28 & Max Eigenvalue Stock Returns\_7 & 0.13 \\ \hline
27 & Resilience\_5 & 0.14 \\ \hline
26 & Mean Closeness Centrality\_1 & 0.14 \\ \hline
25 & Max Eigenvalue Stock Returns\_6 & 0.14 \\ \hline
24 & Largest Component\_1 & 0.14 \\ \hline
23 & 90th Percentile Degree\_2 & 0.14 \\ \hline
22 & Max Eigenvalue Stock Returns\_5 & 0.14 \\ \hline
21 & Resilience\_4 & 0.15 \\ \hline
20 & Max Eigenvalue Stock Returns\_4 & 0.15 \\ \hline
19 & 90th Percentile Degree\_1 & 0.15 \\ \hline
18 & Max Eigenvalue Stock Returns\_3 & 0.16 \\ \hline
17 & Resilience\_3 & 0.17 \\ \hline
16 & Max Eigenvalue Stock Returns\_2 & 0.17 \\ \hline
15 & Max Eigenvalue Stock Returns\_1 & 0.18 \\ \hline
14 & Resilience\_2 & 0.18 \\ \hline
13 & Resilience\_1 & 0.20 \\ \hline
1 to 12  & Log Return\_1 to Log Return\_12 & 0.95--0.44 \\ \hline
\end{tabular}
\caption{42 Selected variables over 63 for the training of the short time period: we don't show the ones with a correlation coefficient less than 0.10.}
\label{tab:selecvariableshort}
\end{table}

\begin{figure}
    \centering
    \includegraphics[width=0.99\linewidth]{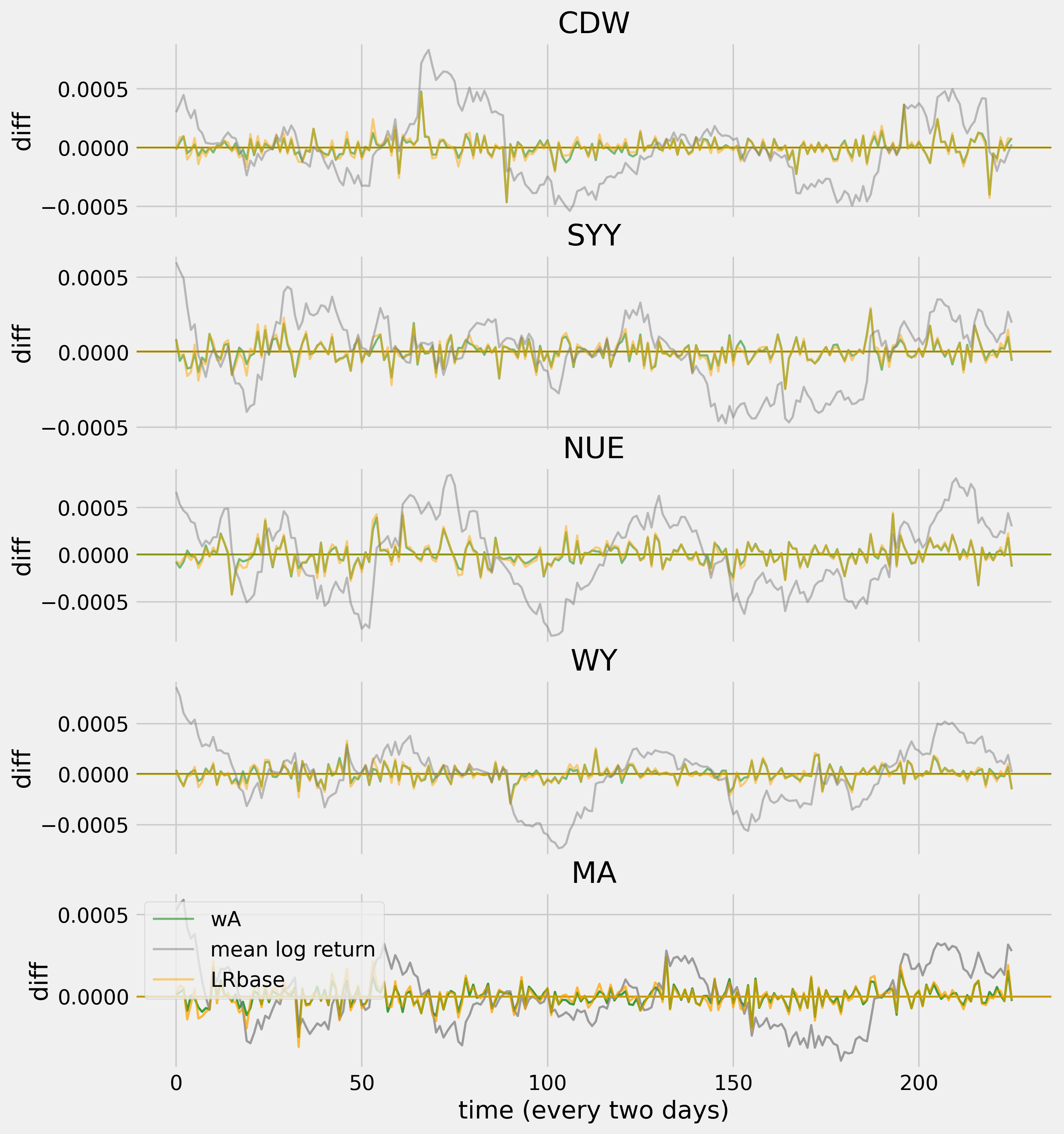}
    \caption{Differences between the wA prediction and the log return (green curves), the LRbase and the log return (orange curves) and the mean of the log return with itself (grey curves) for 5 randomly selected stocks from the testing set. The horizontal lines correspond to the mean of the differences.}
    \label{fig:difflrhours}
\end{figure}

In Fig.\ref{fig:difflrhours} we show the difference of the predicted log return for the wA, the LRbase and the mean of the log return for 5 selected stock in the testing set. There is no qualitative difference between the LRbase prediction and the wA one. Both outperform a simple mean average of the log return.

%% If you have bibdatabase file and want bibtex to generate the
%% bibitems, please use
%%
\section*{References}

\bibliography{networkfinance.bib}

% Make sure to specify the bibliography style

\bibliographystyle{iopart-num}

\end{document}